\documentclass{emulateapj}
\usepackage{natbib}
\usepackage{multirow}

\newcommand{\kev}{keV}
\newcommand{\rxte}{\textit{RXTE}}
\newcommand{\nicer}{\textit{NICER}}
\newcommand{\astroh}{\textit{Hitomi}}
\newcommand{\athena}{\textit{Athena}}
\newcommand{\loft}{\textit{LOFT}}
\newcommand{\fe}{Fe~K$\alpha$}

\newcommand{\eighteen}{4U~1820-30}
\newcommand{\fouru}{4U~1636-536}
\newcommand{\oheight}{4U~1608-522}
\newcommand{\rct}{$\chi_\nu^2$}

\shorttitle{Accretion disk signatures in X-ray Bursts}
\shortauthors{Keek, Wolf, \& Ballantyne}

\begin{document}

\title{Accretion disk signatures in Type I X-ray Bursts: prospects for future missions}

\author{L. Keek,\altaffilmark{1,2} Z. Wolf,\altaffilmark{3} and D. R. Ballantyne\altaffilmark{3}}
\email{laurens.keek@nasa.gov}
\altaffiltext{1}{CRESST and X-ray Astrophysics Laboratory NASA/GSFC, Greenbelt, MD 20771}
\altaffiltext{2}{Department of Astronomy, University of Maryland, College Park, MD 20742}
\altaffiltext{3}{Center for Relativistic Astrophysics, School of Physics, Georgia Institute of Technology, 837 State Street, Atlanta, GA 30332-0430}

\begin{abstract} 
Type I X-ray bursts and superbursts from accreting neutron stars illuminate the accretion disk and produce a reflection signal that evolves as the burst fades. Examining the evolution of reflection features in the spectra will give insight into the burst-disk interaction, a potentially powerful probe of accretion disk physics. At present, reflection has been observed during only two bursts of exceptional duration. We investigate the detectability of reflection signatures with four of the latest well-studied X-ray observatory concepts: \astroh, \nicer, \athena, and \loft. Burst spectra are modeled for different values for the flux, temperature, and the disk ionization parameter, which are representative for most known bursts and sources. The effective area and through-put of a \astroh-like telescope are insufficient for characterizing burst reflection features. \nicer\ and \athena\ will detect reflection signatures in Type I bursts with peak fluxes $\ga 10^{-7.5}$~erg~cm$^{-2}$~s$^{-1}$, and also effectively constrain the reflection parameters for bright bursts with fluxes of $\sim 10^{-7}$~erg~cm$^{-2}$~s$^{-1}$ in exposures of several seconds. Thus, these observatories will provide crucial new insight into the interaction of accretion flows and X-ray bursts. For sources with low line-of-sight absorption, the wide band-pass of these instruments allows for the detection of soft X-ray reflection features, which are sensitive to the disk metallicity and density. The large collecting area that is part of the \loft\ design would revolutionize the field by tracing the evolution of the accretion geometry in detail throughout short bursts.
\end{abstract}


\section{Introduction}
\label{sec:Introduction} 

In low-mass X-ray binaries (LMXBs) that
host a neutron star, the donor star may fill its Roche lobe, and transfer
material via an accretion disk to the neutron star. Hydrogen and helium
accumulated on the neutron star surface can undergo runaway thermonuclear
burning, powering a brief ($10-100\,\mathrm{s}$) Type I X-ray burst
\citep{Grindlay1976,1976Belian,Woosley1976,Maraschi1977,Lamb1978}.
Recurring on typical timescales of hours to days, these bursts are
the most frequent thermonuclear flashes in nature \citep[e.g.,][]{Lewin1993,Strohmayer2006,Galloway2008catalog}.
A bright X-ray burst briefly outshines all other X-ray emitting regions in
the system. 

The X-ray spectrum from the neutron star during the burst is close to
a blackbody (\citealt{swank1977,Paradijs1986spectra}; see also, e.g., \citealt{Suleimanov2010}), and
this spectrum may be reprocessed (or, reflected)
by and scattered off the surrounding disk and the companion star. UV and optical
reprocessing is thought to originate predominantly from the outer regions of the disk
and the companion star \citep{Hynes2006,Paul2012}, whereas X-ray
reflection occurs mostly off the inner disk (\citealt{Ballantyne2004,Keek2014sb2};
see also \citealt{Day1991}) which is struck by a particularly strong ionizing radiation field. As a result, an \fe\ emission line may be visible
in the spectrum near $6.4\,\mathrm{keV}$ as well as an Fe absorption
edge at slightly higher energies, the properties of which depend on
the ionization state of the inner disk. Further spectral features
produced by X-ray reflection include a multitude of lines and a bremsstrahlung
continuum below $\sim1\,\mathrm{keV}$ \citep{Ballantyne2004models}.
When originating from the inner disk, the shape of these spectral features is modified by rotational Doppler
broadening and gravitational redshifting \citep[e.g.,][]{Fabian1989}.
The strength of these effects depends on the distance from the neutron
star, and, therefore, the location of the reflection site can be measured. In addition, the magnitude of the reflection signal encodes information about the disk geometry \citep{Ballantyne2004,He2015}. Therefore, the reflection features in a burst spectrum can potentially reveal a treasure of information on the properties of the accretion environment, similar to how reflection spectroscopy has been invaluable for the study of accretion onto black holes in Active Galactic Nuclei (AGN) and compact binaries \citep[e.g.,][]{Fabian10,Miller2007}.

The short duration of X-ray bursts and the fast evolution of their
spectral properties mean that spectra can only be collected in short
time intervals. Around the peak of the burst this is at most a few
seconds. The quality of the spectra is, therefore, rather limited. 
In normal bursts, reflection features have never been clearly detected, and burst
reflection is not distinguished from the directly observed thermal
emission from the neutron star \citep[e.g.,][]{lapidus85mnras,fujimoto88apj}.
The highest quality burst spectra were obtained with the Proportional
Counter Array \citep[PCA;][]{Jahoda2006} on the \emph{Rossi X-Ray
Timing Explorer} \citep[RXTE;][]{Bradt1993} for two so-called ``superbursts''
with durations of several hours \citep{Strohmayer2002,Strohmayer2002a}.
The long duration allowed for more detailed spectral analysis that revealed
reflection features \citep{Ballantyne2004,Keek2014sb2,Keek2015sb},
and in one case an evolving persistent component \citep{Keek2014sb1}.
The reflection features showed that the inner disk is highly ionized
by the burst, and may temporarily be disrupted. These two observations
show that X-ray reflection during bursts is a powerful tool for investigating
the behavior of accretion disks under sudden strong irradiation.

\cite{Ballantyne2005} considered inflow, outflow, and thermodynamic processes during the interaction of an X-ray burst with the surrounding accretion disk, and found that nearly all processes may be relevant. The variety of possible physical effects means that further theoretical studies as well as improved observational constraints are needed to better understand the impact of an X-ray burst on its surroundings.

Aside from a reflection component, the X-ray spectrum additionally includes components for thermal emission
from the accretion disk, Comptonized emission from the accretion disk
corona, as well as thermal emission from a boundary or spreading layer
where freshly accreted material reaches the neutron star surface \citep{Inogamov1999,Revnivtsev2013}.
These components are also present before the burst as ``persistent''
emission. Typically, their spectral properties are measured at that
time, and one assumes them to remain unchanged during the burst. All
of the burst spectrum in excess of the combined persistent parts is
assumed to be blackbody emission from the neutron star \citep[see also][]{2002Kuulkers}.
However, there is now evidence that the persistent emission is being altered by the burst, as its normalization may increase \citep{Worpel2013,Zand2013,Keek2014sb1,Worpel2015,Degenaar2016}, possibly due to Poynting-Robertson drag \citep{Walker1992}, and its spectrum
appears to soften \citep{Maccarone2003,Chen2012,Chen2013,Ji2014,Keek2014sb1}. Therefore, when studying burst reflection, care must be taken to distinguish reflection from an evolving persistent spectral component.

In this paper we investigate the capabilities of different instrumentation
for detecting burst reflection. We simulate a wide range of burst observations with
four X-ray observatories: \nicer, \astroh, \athena, and \loft\ 
(Section~\ref{sec:Method}). Due to their large collecting areas,
\nicer, \athena, and \loft\ will be able to detect reflection features
in bright bursts, even if their duration is short (Section~\ref{sect:res}).
We discuss how burst reflection can be applied to study the accretion environment of neutron stars in LMXBs, 
and how it impacts neutron star science such as constraints on the dense matter equation of state (Section~\ref{sect:discuss}).
We conclude that X-ray burst reflection will become an important method
to study the properties and behavior of accretion disks in LMXBs.
(Section~\ref{sect:conclusions}).

\section{Method}
\label{sec:Method}

\subsection{Instrumentation}
\label{sec:Instrumentation}

The detection of X-ray burst reflection is highly dependent on the properties of the instrumentation, such as the energy band, the effective area, and the spectral resolution. We study the detectability of reflection using the expected response of four X-ray observatories that are either recently launched, being constructed, or planned for a launch opportunity further in the future.

The {\it Neutron Star Interior Composition Explorer} \citep[NICER,][]{Gendreau2012NICER} is planned for launch in 2016, and will be placed on the International Space Station. Its X-ray Timing Instrument (XTI) consists of 56 units, each with an X-ray concentrator optic and a silicon strip detector. It is not an imaging instrument, but it collects all photons from the $30\ \mathrm{arcmin}^2$ field of view. Its detectors are sensitive in the $0.2-10\ \mathrm{keV}$ energy band, and its effective area is approximately $2000\ \mathrm{cm^2}$ at $1\ \mathrm{keV}$. By using of a large number of detectors, \nicer\ can handle high photon count rates.

\athena\ \citep{BarconsATHENA} is a proposed large mission for the European Space Agency to be launched in the late 2020s. It will combine a collecting area larger than \nicer\ with imaging capabilities and high spectral resolution. We consider the imaging CCD detector: the Wide Field Imager \citep[WFI,][]{Meidinger2014AthenaWFI}. It is sensitive in the $0.3-12\ \mathrm{keV}$ band, and has a $\sim 17,000\ \mathrm{cm^2}$ effective area around $1\ \mathrm{keV}$.

Beyond {\athena}, we may speculate about a future observatory
with an even larger detector area. The \emph{Large Observatory For
X-ray Timing} \citep[LOFT,][]{Feroci2014LOFT} was a mission concept
proposed to ESA. Although it was not selected, parts of its design may be incorporated in other missions. With an effective area of $8.5\,\mathrm{m^{2}}$
combined with high throughput, it promised unprecedented opportunities
for studying short transient events such as Type I X-ray bursts \citep{Zand2015LOFT}.
We use the M4 configuration of its Large Area Detector (LAD). Contrary to the other three instruments that we consider, the LAD is not
sensitive below $2\ \mathrm{keV}$: its energy band is $2-80\ \mathrm{keV}$.

\astroh\ \citep[named {\it ASTRO-H} pre-launch,][]{TakahashiASTROH} was launched in February 2016 and operated until the end of March. As it had significantly different capabilities than the other instruments considered here, it is valuable to include it in the present study. \astroh\ hosted several instruments, and we consider the Soft X-ray Imager \citep[SXI,][]{Hayashida2014astrohSXI}. SXI was a CCD imaging detector sensitive in the $0.4-12\ \mathrm{keV}$ energy band, and at $1\ \mathrm{keV}$ the effective area was $\sim 590\ \mathrm{cm^2}$. Compared to \nicer's detectors, the imaging capability allows for a lower background, but it reduces the maximum photon count rate that the instrument can process. In Section\ \ref{sec:Pileup} we will discuss the impact of this on the ability to detect reflection.

\begin{table*}
\centering
	\caption{Properties of the simulated instruments, as well as the employed response and background files.} 
    \label{table:missions}
   \begin{tabular}{cccccccc}
\hline
Mission & Instrument & Bandpass & $6$~keV Area & Background File & RMF & ARF & Reference \\ 
 & & (keV) & (cm$^{2}$) & & & & \\ \hline
\nicer\ & XTI & $0.2$--$10$ & 600 & 15174.bkg & nicer\_xti.rmf & --- & \citet{Gendreau2012NICER} \\
\astroh\ & SXI & $0.4$--$12$ & 360 & astroh\_bkgnd.pi & astroh.rmf & astroh\_pointsource.arf & \citet{TakahashiASTROH} \\
\athena\ & WFI & $0.3$--$12$ & 2500 & athena\_wfi\_1190\_bkgd  &  athena\_wfi\_1190 & --- & \citet{BarconsATHENA}\\
 & & & & \_sum\_psf\_onaxis\_w & \_onaxis\_w\_filter &  & \\
 & & & & \_filter\_20150327.pha & \_v20150326.rsp &  & \\
\loft\ & LAD & $2.0$--$80$ & 81700 & LAD\_M4\_v2.0.bkg & LAD\_M4\_v2.0.rmf & LAD\_M4\_v2.0.arf & \citet{Feroci2014LOFT} \\
	\hline
  
   \end{tabular}
\end{table*}

\begin{figure}
\includegraphics{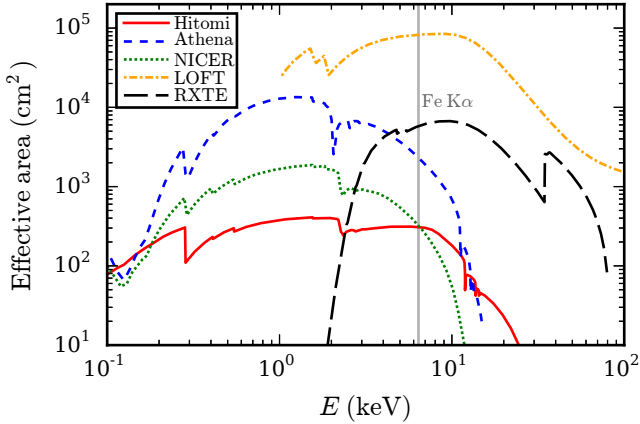}

\protect\caption{\label{fig:area} The effective area as a function of energy, $E$, for the instruments considered in this paper including \rxte/PCA (Sect.\ref{sub:rxte}). The vertical line indicates the location of the \fe\ line near $6.4$ \kev. \loft\ has a larger area than \rxte\ (5 PCUs) at this energy. However, the effective areas of \astroh, \nicer, and \athena\ extend to much lower energies than \rxte, allowing for additional reflection features to be detected.}
\end{figure}

We simulate burst reflection spectra for all missions using the response matrix files (RMFs), ancillary response files (ARFs), and background files that have been made available by the instrument teams. Table~\ref{table:missions} lists which exact files are employed, and Figure~\ref{fig:area} compares the effective areas. The figure also includes the area of \rxte/PCA (assuming all 5 Proportional Counter Units [PCUs] are operating), the only instrument to have successfully detected burst reflection. A more detailed comparison to \rxte\ is made in Sect.~\ref{sub:rxte}.

\subsection{Spectral Model}
\label{sec:Model}

\begin{figure}
\includegraphics{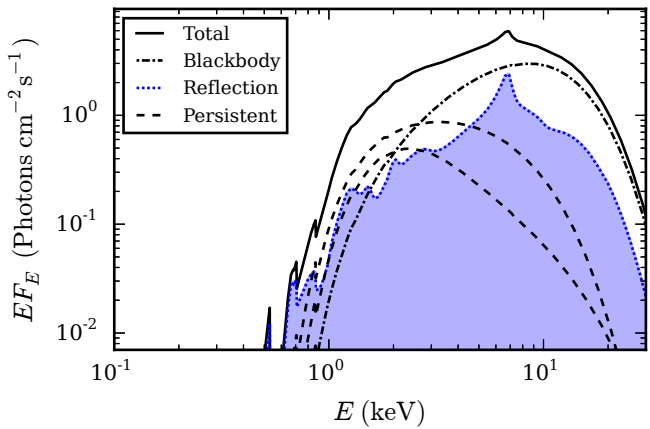}

\protect\caption{\label{fig:model} Illustration of our spectral model and its components for $kT=3.0\ \mathrm{keV}$ and $\log\xi=3.0$ (Table~\ref{table:RefParams}, Section~\ref{sec:Model}): energy flux, $EF_E$, as a function of the spectral energy, $E$. Reflection (shaded area) produces a strong Fe~K$\alpha$ line near $6.4$~keV. The persistent emission is modeled with two components.}
\end{figure}

Three distinct emission components can contribute to the X-ray spectrum during a burst: the persistent emission from the underlying accretion disk, the thermal blackbody emission from the NS surface, and the reflection of that blackbody from the disk surface. Additionally, a thin boundary or spreading layer may  be present, which during a burst may cover a substantial part of the neutron star \citep{Inogamov1999,Revnivtsev2013}. We do not explicitly include a spreading layer, but our setup is equivalent to complete coverage by this layer. We aim for our spectral model to be representative of a wide range of bursts, covering the majority of the known bursting sources. We employ XSPEC version 12.8.2 \citep{Arnaud1996} to create a spectral model that includes all three components, along with interstellar absorption (Figure~\ref{fig:model}). Using XSPEC terminology, the model reads as \texttt{phabs*(constant*(diskbb + compTT) + rdblur*cflux*atable(bbrefl\_1xsolar\_0-5r.fits))} with the parameter values shown in Table~\ref{table:RefParams}. Below we describe the components in detail.

We employ a blackbody component with temperature $kT$ as well as reflection of this blackbody off a photoionized accretion disk \citep{Ballantyne2004models}. The reflection spectrum is sensitive to the ionization state of the illuminated layer and, as this is a constant density model, can be parameterized using the ionization parameter $\xi=4\pi F_{x}/n_{\mathrm{H}}$, where $F_{x}$ is the irradiating flux at the surface of the reflecting slab and $n_{\mathrm{H}}$ is its density. Both $kT$ and $\xi$ evolve during a burst \citep{Ballantyne2004,Keek2014sb2}; therefore, we consider a range of values for these parameters in order to determine how they may affect the detectability of the reflection signal. The reflection fraction, $R$, is defined as the flux ratio of the reflection and the blackbody parts. We use $R=0.5$, which is approximately the value predicted for a flat accretion disk viewed under a small inclination \citep[$\sim 30^\circ$; e.g.,][]{He2015}, and which is a typical value found in the superburst analyses \citep{Ballantyne2004,Keek2014sb2}. The composition of the accretion disk is assumed to be solar, consistent with most bursters. A subset of bursters are so-called Ultra-Compact X-ray Binaries (UCXBs), where the accretion composition is thought to be hydrogen-deficient \citep[e.g.,][]{intZand2007}. However, because the metal content for UCXBs is likely similar to solar, and metals produce the prominent reflection features such as the \fe\ line, our conclusions on the detectability of reflection will also be applicable to these sources. A \texttt{cflux} component operating on the combined blackbody and reflection model sets the unabsorbed bolometric flux of the burst, and allows for a range of burst fluxes to be simulated.

The reflection signal may arise from the inner regions of the accretion disk and will be sculpted by relativistic effects \citep[e.g.,][]{Fabian1989,Ballantyne2004}. Thus, the reflection model is convolved with the \texttt{rdblur} relativistic blurring kernel that accounts for the effects of the emission being released in a Schwarzschild metric. As we are simply interested in the detectability of the reflection component, the \texttt{rdblur} parameters are kept fixed throughout the experiment at values indicative of reflection from the inner disk, as seen from \eighteen\ \citep{Ballantyne2004}, assuming a typical inclination angle of the disk with respect to the line of sight of $30^\circ$. Although the location of the reflection signal may evolve throughout a burst, these changes will likely be undetectable for most of the considered instruments unless they arise from a superburst.

The persistent emission of LMXBs depends on their spectral state \citep[e.g.,][]{1989Hasinger}. It is thought that in the soft state, the inner disk extends closest to the neutron star, which maximizes the reflection signal. Furthermore, the soft state is associated with higher persistent flux and a higher burst frequency. Therefore, we use a soft persistent spectrum in our model, but note that reflection will be even better detectable in the hard state when the persistent flux is lower (for equal reflection fraction). As a typical example of a soft spectrum, we employ the spectral shape measured in broad-band {\it BeppoSAX} observations of the transient source \oheight\ during outburst \citep{Keek2008}. This includes thermal emission from the disk (\texttt{diskbb}) plus a Comptonized component (\texttt{compTT}). When in outburst, \oheight\ exhibits frequent bright bursts with peak fluxes of $F\simeq10^{-7}$~erg~cm$^{-2}$~s$^{-1}$, making it a prime candidate for detecting burst reflection. Our results will not be strongly dependent on the details of the persistent spectrum, because we test how well reflection can be distinguished from the persistent components by fitting for a multiplicative factor for the latter.

The multiplicative factor, denoted $f$, in front of the persistent emission (\texttt{constant} in XSPEC) may also represent the possibility of a change in the persistent emission during the burst. Such behavior has been inferred for a variety of Type~I bursts \citep{Worpel2013,Worpel2015} and measured in the 2001 superburst from \fouru\ \citep{Keek2014sb1}. The exact interpretation of this increase in persistent emission is not understood, and it may be related to an increase of accretion onto to the neutron star or to changes in the disk corona due to the influence of the burst. This parameter is set to unity for a burst flux of $F=10^{-7}$~erg~cm$^{-2}$~s$^{-1}$ (typical peak flux of a bright burst), and is scaled appropriately as the flux is changed so that it remains the same fraction of the total flux.

Photoelectric interstellar absorption is taken into account using the \texttt{phabs} model with cross-sections from \cite{Balucinska1992} and abundances from \cite{Anders1989}. Most X-ray bursting sources are near the Galactic Center, where absorption is relatively strong ($N_\mathrm{H}\sim 10^{22}\ \mathrm{cm^{-2}}$). For consistency, we use the absorption column of $N_\mathrm{H}=0.891\times 10^{22}\ \mathrm{cm^{-2}}$ measured for \oheight\ during the same observation from which our persistent spectrum originates. Additionally, in Section~\ref{sec:low_abs} we investigate the case of a smaller absorption column.

\begin{table}
  \caption{Parameters of the simulated Type I X-ray burst spectra.}
 \label{table:RefParams}
  \begin{tabular}{cccc}
    \hline
    Model Component& Parameter & Units & Value$^a$\\ \hline	\hline
    phabs & nH & $10^{22}$~cm$^{-2}$& 0.891\\     \hline
    constant & $f$ & & $1^b$ \\     \hline
    \multirow{2}{*}{diskbb}	& $kT_{\mathrm{in}}$ & keV & 2.38 \\
    							& norm & $R_{\mathrm{km}}^2/D_{10}^2$ & 18.3 \\     \hline    							
  	\multirow{5}{*}{compTT}	& $kT_0$ & keV & 0.478 \\
    						& $kT_e$& keV & 3.6 \\
    						& $\tau$ & & 3.7 \\
    						& approx. && 1 \\
    						& Norm.& & 0.4 \\    \hline
       \multirow{4}{*}{rdblur} & Emissivity & & -2 \\
    						& $r_{\mathrm{in}}$ & $GM/c^2$ & 20 \\
    						& $r_{\mathrm{out}}$ & $GM/c^2$ & 1000 \\
    						& Incl. Angle & degrees & 30 \\    \hline 
    \multirow{3}{*}{cflux} & $E_{\mathrm{min}}$ & keV & 0.001 \\
    						& $E_{\mathrm{max}}$ & keV & 100.0 \\
    						& $F$ & erg cm$^{-2}$~s$^{-1}$ & $10^{-8}$--$10^{-4.5}$ \\ 	\hline
               Blackbody +  & $\log \xi$ & $\log$(erg cm s$^{-1})$ & $1.5$--$3.5$ \\
			Reflection		& $kT$ & keV & $1.5$--$3.0$ \\
							& $R$ & & 0.5 \\ \hline
  \end{tabular}

$^a$ Listed are either the fixed value of the parameter or the range over which it is varied.

$^b$ Value at $\log F=-7$, but scaled appropriately for other values of $F$ so that the relative strength of persistent emission is unchanged.
\end{table}

\subsection{Spectral Simulations}
\label{sec:Calcs}

Using the spectral model (Table~\ref{table:RefParams}) and the responses of the instruments considered here (Table~\ref{table:missions}) we create a set of simulated spectra. The instrumental background is included as well as statistical fluctuations following a Poisson distribution. Neighboring spectral bins with fewer than 20 counts are grouped to ensure $\chi^2$ statistics are applicable. In order to cover the wide variety of known bursts and sources,  we investigate a broad range of values for the burst flux $F$, the burst temperature $kT$, and the ionization parameter of the reflector $\log \xi$. Therefore, for each instrument, the spectral model is simulated over a three-dimensional space defined by these three parameters.

The range of values for temperature, $1.5\leq kT \leq 3.0$, is resolved in steps of 0.5, and covers the peak of the hottest bursts as well as the tails of weaker bursts \citep[e.g.,][]{Galloway2008catalog}. The strength of the reflection features are a strong function of $\xi$ and $kT$ \citep{Ballantyne2004models}. Therefore, we study a range of values for the ionization parameter, $1.5\leq \log \xi\leq 3.5$, again resolved in steps of 0.5, which includes the values observed during superbursts \citep{Ballantyne2004,Keek2014sb2}.

The statistical quality of the spectra depends on the observed number of counts, which is proportional to the product of the flux and the exposure time. For simplicity, the exposure time of each simulation is maintained at one second. This is approximately the time resolution required for time-resolved spectral analysis of the shortest bursts. For longer bursts, especially during the tail when the spectral parameters evolve more slowly, longer exposures can in principle be taken. Results for longer exposure times can be found by appropriately scaling the burst flux, such that the same number of counts is obtained. For the bolometric burst flux we use the values $-8.0\leq \log F \leq -4.5$, with steps of $0.5$. This range includes $\log F \approx -7.0$ for the typical brightest known bursts \citep[e.g.,][]{Galloway2008catalog} as well as an order of magnitude lower fluxes. Furthermore, for the two mentioned superburst observations spectra were collected over $64$~s intervals \citep{Ballantyne2004,2004Kuulkers}. This is equivalent to looking for 1 second at a burst that is $64$ times brighter, and for the brightest bursts this extends the flux range to $\log(64\times 10^{-7}) = -5.2$. Spectra should not be accumulated over a longer duration because the spectrum changes as the temperature of the neutron star changes significantly, but to be inclusive, we take one extra step of 0.5, and include a flux of $\log F=-4.5$.

These choices of parameter values span the full relevant range of photospheric temperatures observed from all known bursting sources, as well as the full range of expected ionization states of the accretion disk. Moreover, $F$ not only represents the intrinsic burst flux, but can be scaled to match different exposure times and source distances.

\section{Results}
\label{sect:res}

\subsection{Detectability of Reflection}
\label{sub:detect}

To quantify the detectability of burst reflection, we investigate whether the reflection parameters can be retrieved from the simulated data, or whether a blackbody alone adequately describes the data. Each simulated spectrum is fitted twice for each instrument: once with the same reflection model used to generate the simulation (with $f$, $kT$, $R$, $\log F$ and $\log \xi$ as fit parameters) and once with a standard blackbody model. The latter model replaces the reflection component with a simple blackbody, and now $f$, $kT$, and the blackbody normalization are the fit parameters. All other parameters are kept fixed. The reduced $\chi^2$, \rct, of each fit is recorded along with the best-fit values of the parameters, including their $1\sigma$ uncertainties.

A quick overview of the results is shown in Table \ref{table:RefMinMax}, where the {\rct} values (averaged over all $\log \xi$ and $\log F$ values) are listed for each $kT$ and for each instrument (a comparison to \rxte\ follows in Section~\ref{sub:rxte}). Unsurprisingly, the reflection model is able to provide a good description of the simulated spectra with little scatter in the \rct\ values. However, the blackbody model --- which is commonly used in burst analyses --- may produce very poor fits to the simulated observations from \nicer, \athena, and \loft. The \rct\ values also appear to become progressively worse as $kT$ increases, as the peak of the blackbody spectrum moves to the edge of the bandpass. A \astroh-like mission, on the other hand, will have difficulty to distinguish the reflection features in the burst spectra because of its smaller effective area.

\begin{table}
  \caption{The average \rct\ of fits to reflection spectra.}
 \label{table:RefMinMax}
 \begin{center}
  \begin{tabular}{ccccc}
    \hline
    $kT$ & \astroh\ & \nicer\ & \athena\ & \loft\  \\ \hline
	\multicolumn{5}{c}{Reflection Model $\bar{\chi^2_{\nu}}$~$^a$} \\
    1.5 & 0.96 $\pm$ 0.10 & 0.96 $\pm$ 0.08 & 0.97 $\pm$ 0.06& 0.99 $\pm$ 0.08 \\ 
    2.0 & 0.96 $\pm$ 0.14 & 0.98 $\pm$ 0.08 & 0.99 $\pm$ 0.06& 0.98 $\pm$ 0.06 \\
    2.5 & 0.96 $\pm$ 0.13 & 0.96 $\pm$ 0.10 & 0.99 $\pm$ 0.06& 1.01 $\pm$ 0.07 \\
    3.0 & 1.03 $\pm$ 0.23 & 0.95 $\pm$  0.11 &0.97 $\pm$ 0.06& 0.99 $\pm$ 0.08 \\ \hline
        \multicolumn{5}{c}{Blackbody Model $\bar{\chi^2_{\nu}}$} \\
    1.5 & 1.03 $\pm$ 0.18 & 1.76  $\pm$ 1.76& 5.39 $\pm$ 9.48 & 59.21 $\pm$ 133.24\\
    2.0 & 1.03 $\pm$ 0.22 & 1.85 $\pm$ 1.88 & 5.54 $\pm$ 9.78 & 80.35 $\pm$ 191.43\\
    2.5 & 1.04 $\pm$ 0.23 & 2.00 $\pm$ 2.25 & 5.82 $\pm$ 10.48& 80.28 $\pm$ 193.25 \\
    3.0 & 1.09 $\pm$ 0.25 & 2.18 $\pm$ 2.70 & 6.39 $\pm$ 11.84& 71.06 $\pm$ 173.71\\ \hline
  \end{tabular}
 \end{center}
 
 $^a$ The average \rct\ and its standard deviation at each $kT$ for \astroh, \nicer, \athena, and \loft. The average is taken across all $\log \xi$ and $\log F$ values. Results are presented both for fits with the same reflection model that was used to simulate the spectra (top) and for fits with a model that does not include a reflection spectrum (bottom).
\end{table}

\begin{figure*}
\centering
\includegraphics[width=0.95\textwidth]{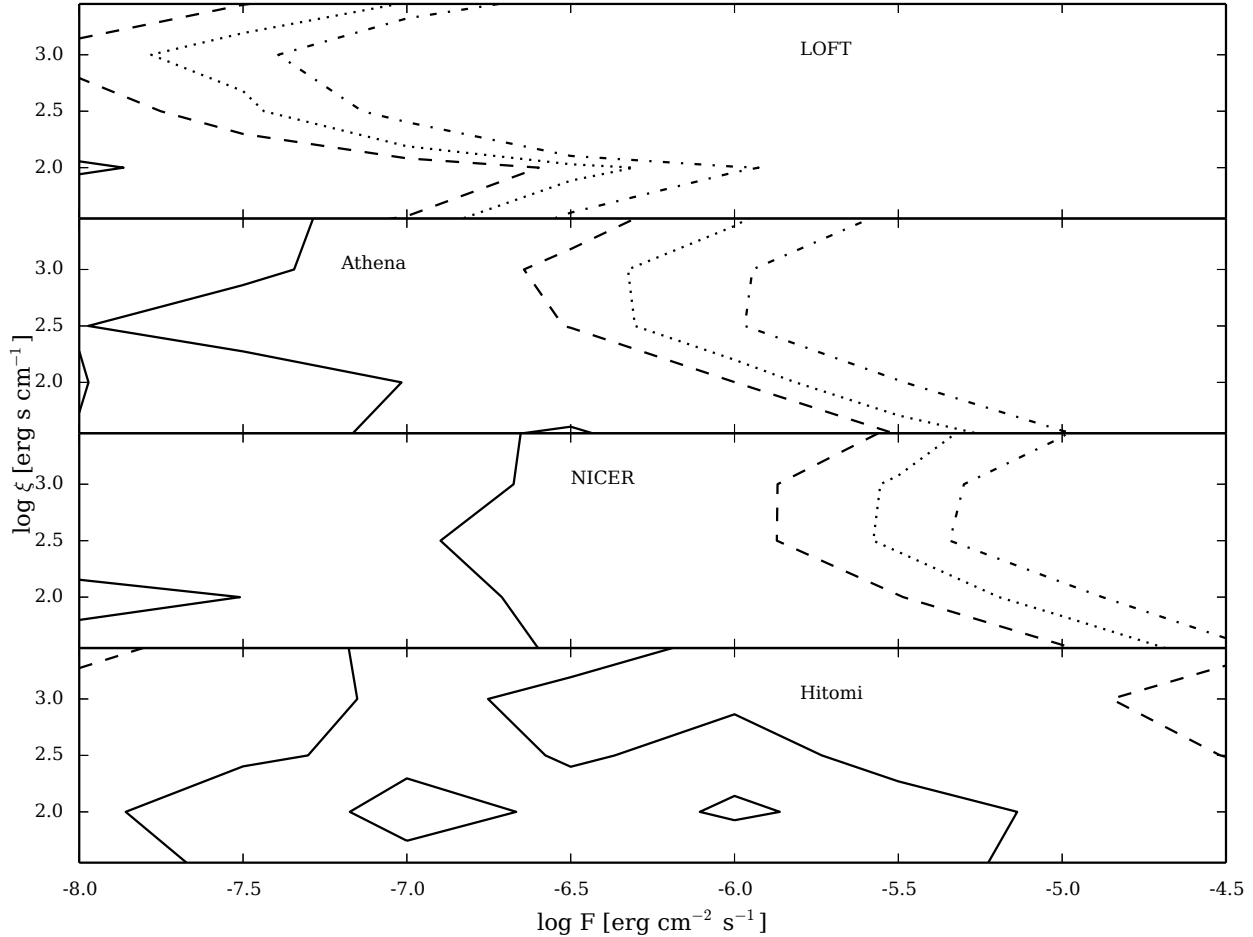} 
\caption{Contours of goodness of fit, \rct, from fitting the blackbody model to simulated burst reflection spectra for four instruments. Shown are simulations for a range of values for the ionization parameter, $\log\xi$, and the unabsorbed bolometric burst flux, $F$. A blackbody temperature of $kT=3$~keV is used (the results at other temperatures are qualitatively similar). The simulations have an exposure time of $1\ \mathrm{s}$, and a high $F$ is equivalent to a longer exposure at correspondingly lower $F$. The contour levels are {\rct}=1.0 (solid), 1.5 (dashed), 2.0 (dotted), and 3.0 (dashed-dotted).}
\label{fig:chicompare}
\end{figure*}

The standard deviations of the average \rct\ shown in the lower half of Table~\ref{table:RefMinMax} are similar in magnitude to the average \rct\ values themselves and increase with $kT$. Therefore, there is clearly a wide range of fit results obtained within the parameter space defined by $(kT, \log \xi, \log F)$. A more detailed view of the variety of $\chi^2_\nu$ obtained when applying the blackbody model to the simulated data is presented in Figure \ref{fig:chicompare}, where contours of {\rct} are shown as functions of $\log F$ and $\log \xi$ at $kT=3$~keV for each instrument considered (the results are qualitatively similar at other blackbody temperatures). 

\begin{figure}
\includegraphics{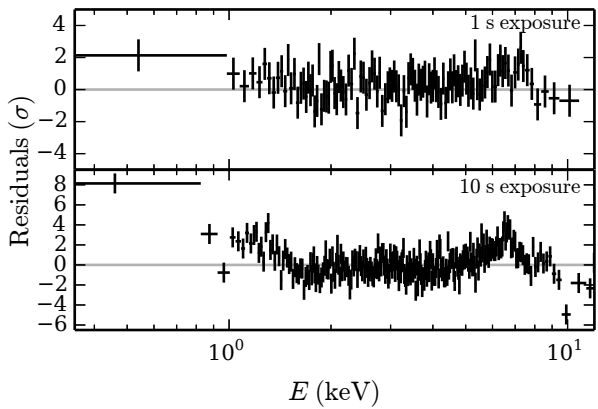}
\protect\caption{\label{fig:nicer_residuals}Residuals of blackbody fits to
two \nicer\ reflection spectra (in units of the $1\sigma$ uncertainty of the data points). The spectra are simulated for $kT=3\,\mathrm{keV}$,
$\log\xi=3$, and $\log F=-7$. We show a $1\,\mathrm{s}$ (top) and
a $10\,\mathrm{s}$ exposure (bottom). The iron line and the soft excess are clearly
visible.}
\end{figure}
The effective area of a \astroh-like telescope is relatively small, such that reflection signatures in Type~I X-ray bursts are unlikely to be strongly detected by this mission unless it was a burst of historic brightness or duration (see also Section~\ref{sec:Pileup}). In contrast, the \nicer\ reflection spectra significantly deviate from a blackbody for $\log F \ga -6.0$. The large effective area of \athena\ could enable the detection of reflection signatures at even lower flux levels. However, it may not be able to handle the large photon counts that an X-ray burst provides (Section~\ref{sec:Pileup}). The \loft\ design includes an even larger collecting area, and the blackbody model is inadequate at all but the lowest considered flux values. \nicer\ and \loft\ will, therefore, be best suited to detect reflection. Figure~\ref{fig:nicer_residuals} illustrates the deviations of a \nicer\ reflection spectrum from a blackbody. Clearly visible are the Fe~K$\alpha$ line near $6.4$~keV as well as a soft excess below $\sim 2$~keV.

The contours of Figure~\ref{fig:chicompare} show for all instruments that reflection signatures are more easily detected for $\log \xi \approx 2.5$. This is the result of two features in the X-ray reflection spectrum \citep{Ballantyne2004models}. First, the equivalent width of the Fe K$\alpha$ line is strongest at this ionization parameter because it is dominated by recombination onto He-like Fe, and the other metals in the irradiated slab (e.g., C, N and O) are highly ionized, leading to less absorption at energies around 6~keV and a more prominent Fe line. In addition, the gas at these ionization parameters is full of hot electrons and ionized metals that together produce a significant bremsstrahlung-dominated soft excess at lower energies. As all the instruments considered here with the exception of \loft\ have a band-pass that extends substantially below $1$~keV (Table~\ref{table:missions}, Figure~\ref{fig:area}), this soft excess provides a significant deviation from the blackbody shape and allows the reflection signal to be more easily detected. At larger values of $\log \xi$, the Fe K$\alpha$ line becomes weaker due to the increased ionization and the soft excess becomes stronger, but the overall detectability of the reflection signal is reduced. For $\log \xi \la 2.5$, both the soft excess and the Fe K$\alpha$ equivalent width are smaller, and it becomes significantly more challenging for the instruments to detect the reflection signal in the burst spectra.

Figure~\ref{fig:chicompare} indicates the region of parameter space where the reflection features are strong enough that a blackbody model cannot provide a good fit to the simulated spectrum. However, the reflection signal remains in the data even if a blackbody model is a good statistical fit to the spectrum. An example of this can be seen in the top panel of Figure~\ref{fig:nicer_residuals}, where we show the residuals of the blackbody fit to a 1~s \nicer\ observation for a burst with $\log F=-7$. According to the contours of Figure~\ref{fig:chicompare}, the blackbody model is an acceptable fit to these data, yet the residuals clearly show the effects of the reflection spectrum with excesses at $\approx 6$~keV and at lower energy. In this case, a fit with the reflection model would likely be a significant improvement over the blackbody model. Therefore, it is interesting to examine our fits results in a new way and determine the region of parameter space where the reflection model is a significant improvement over the blackbody model. 
\begin{figure*}
\centering
\includegraphics[width=0.95\textwidth]{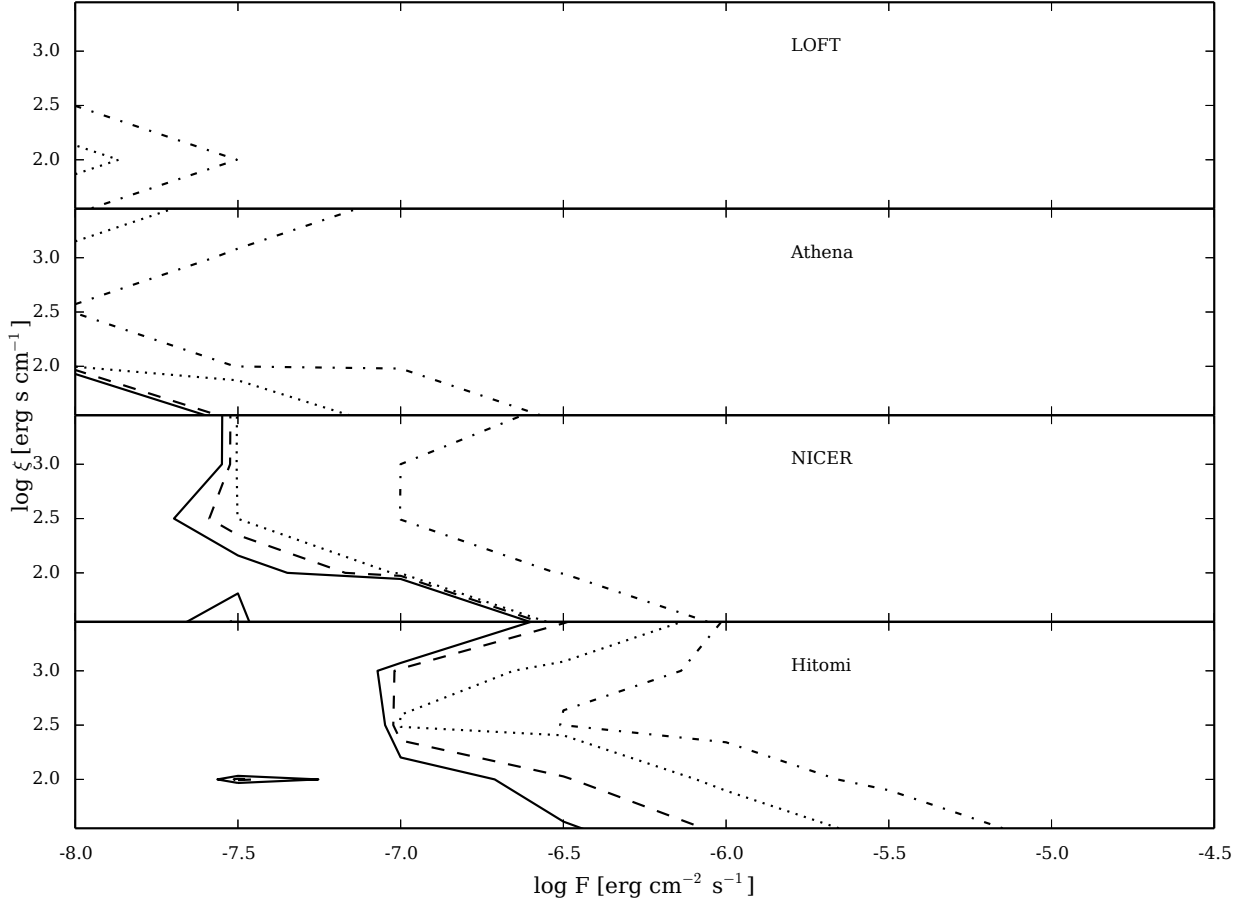} 
\caption{Contours of the F-test probability, illustrating the significance of the improvement in $\chi^2$ when a reflection model is fit to the simulated spectra compared to the simple blackbody model. The results for a blackbody temperature of $kT=3$~keV are shown (the results are qualitatively similar at other temperatures). The contour levels of significance are 90\% (solid), 95\% (dashed), 99\% (dotted), and 99.99\% (dashed-dotted).}
\label{fig:Ftestcompare}
\end{figure*}
To this end, we compute the F-test probability for each spectrum to determine the significance of any improvement the reflection model makes over the blackbody model. Contours at different significance levels are shown in Figure~\ref{fig:Ftestcompare} for the $kT=3$~keV models. These contours indicate which bursts are better described by the reflection model, whereas Figure~\ref{fig:chicompare} shows where the blackbody model cannot adequately describe the spectra (a more stringent requirement). The reflection model is a significant improvement over the blackbody model for $\log F \ga -8$ (\loft, \athena), $-7.5$ (\nicer), $-7$ (\astroh). These fluxes are substantially smaller than the limits needed for the blackbody model to provide a poor description of the data (Figure~\ref{fig:chicompare}), and therefore implies that reflection features will be detectable down to these flux limits, but constraints on the reflection parameters may be relatively poor.

\subsection{Uncertainties in the Measured Parameters}
\label{sub:uncertainty}
\begin{figure}
\centering
\includegraphics[width=0.49\textwidth]{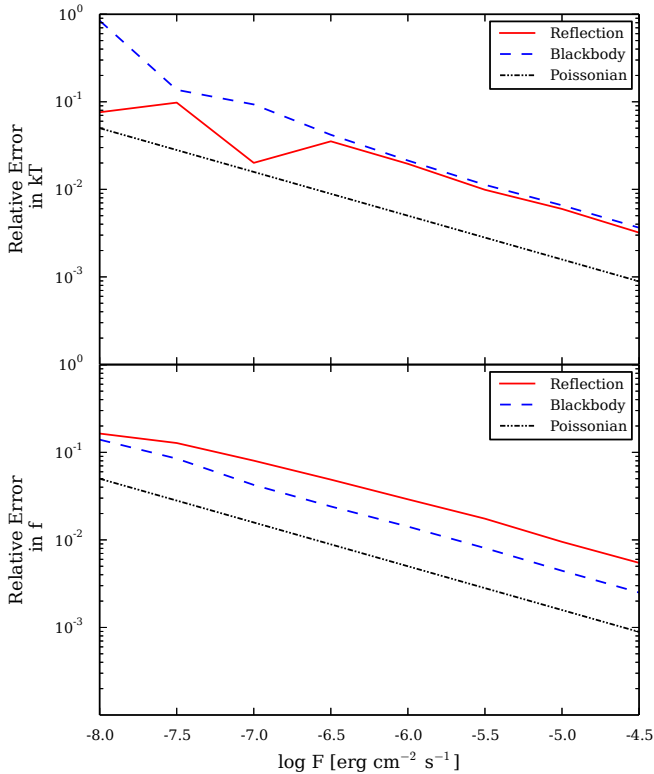}
\caption{(Top) Relative error in measurements of $kT$ from a Type~I X-ray burst with a reflection component as a function of burst flux $\log F$. The simulated X-ray spectra assume observations by \nicer, $kT=3$~keV and a disk ionization parameter of $\log \xi=3$. Data are fit with models that includes disk reflection (solid line) and one that includes emission from only the blackbody and persistent components (dashed line). The dotted line plots the slope expected for a Poisson distribution to help guide the eye. Deviations from this slope at small $\log F$ are due to statistical fluctuations. Both models constrain $kT$ to similar precision. (Bottom) As above, but now plotting the error in determining $f$, the scale factor of the persistent emission. The extra component in the reflection model leads to additional degeneracies in determining the normalization of the spectra and thus gives a larger error in $f$.}

\label{fig:uncstack}
\end{figure}

Many spectral analyses of Type~I X-ray bursts omit the possibility of a reflection component in the model. It is therefore interesting to consider if this omission has an effect on the uncertainty in the spectral parameters. We plot in Figure~\ref{fig:uncstack} the relative uncertainty in two crucial X-ray burst parameters ($kT$ and $f$) as a function of flux for simulated \nicer\ bursts. The specific reflection model used for the figure has $kT=3$~keV and $\log \xi=3$, although the results are qualitatively similar for other parameters and other observatories. The relative error on the parameters is defined using the $1\sigma$ uncertainty from the spectral fit compared to the true value of the parameter. We compare the errors for fits with the reflection model and with the model that omits reflection.

For both $f$ and $kT$ the measured error decreases with flux as expected for a Poisson distribution. For the case of $kT$, it is independent of the spectral model used to fit the simulated data.\footnote{Deviations from the behavior expected for a Poisson distribution are seen in Figure~\ref{fig:uncstack}, in particular at small $\log F$. At these small fluxes, the statistically random noise added to the simulation is most significant. Thus, these deviations are simply due to the single realization of a spectrum with a significant contribution from Poisson noise.} Thus, the measured relative error of $kT$ appears to be robust to neglecting reflection in the spectral fitting. In contrast, the uncertainty in $f$, which measures the relative strength of the persistent emission, is higher when fit with a model that includes reflection, even though this model was used to generate the simulated data. This is because the reflection model has three parameters that can be adjusted to set the overall normalization of the spectrum ($f$, $\log F$, and the reflection fraction $R$), whereas the blackbody model only has $f$ and $\log F$. This extra degree of freedom in the reflection model leads to a larger uncertainty in the normalization of the relatively weak persistent emission.

\begin{figure}
\centering
\includegraphics[width=0.49\textwidth]{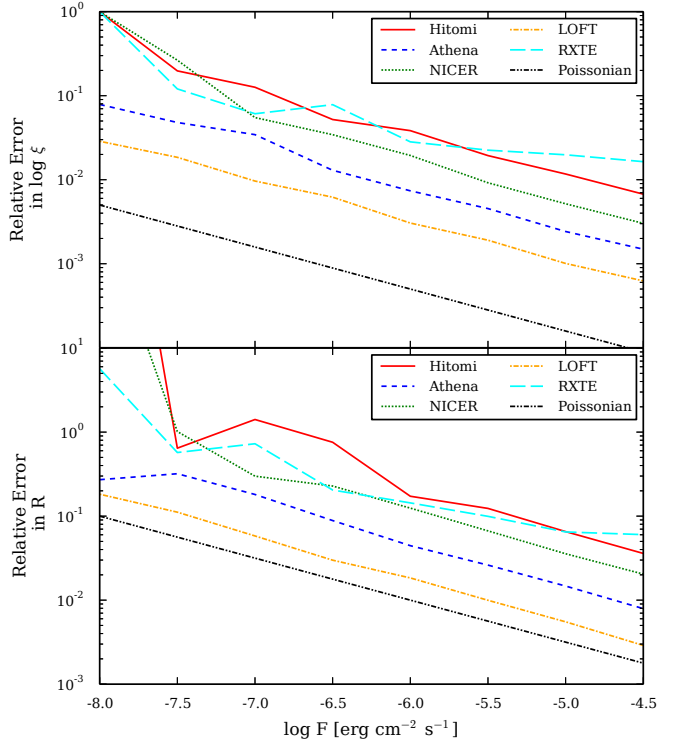}
\caption{(Top) Relative error in measurements of the reflection fraction $R$ from bursts observed by all four facilities as a function of burst flux $\log F$. The simulated X-ray spectra assume $kT=3$~keV and a disk ionization parameter of $\log \xi=3$. The solid line near the bottom of the panel shows the slope for Poisson noise.  (Bottom) As above, but now showing the error in $\log \xi$, the ionization parameter of the reflecting disk.}
\label{fig:reflectuncer}
\end{figure}

Next, we consider the relative uncertainties in the reflection parameters: the reflection fraction $R$ and the ionization parameter $\log \xi$, as a function of flux for a burst with $kT=3$~keV and $\log \xi=3$ as observed by all four observatories (Figure~\ref{fig:reflectuncer}). 
As with the burst parameters, the errors on the parameters increase to lower fluxes due to Poisson noise, but for \nicer\ and \astroh\ the relative errors of both parameters increase sharply at $\log F \le -8$ where the reflection signal is no longer significantly detectable (Figure~\ref{fig:Ftestcompare}). \nicer\ can measure $\log \xi$ with a relative error $<0.1$ for fluxes $\log F \ga -7$. \athena\ and \loft\ measure $\log \xi$ with this relative error for all our modeled bursts, and can even constrain $\log \xi$ with an error $< 0.01$ for $\log F \ga -6$ (\athena) and $\log F \ga -7$ (\loft), respectively. 

As $R$ is related to the covering factor of the irradiated accretion disk, measurements of $R$ will provide information on the disk geometry.  Constraining $R$ is a more difficult challenge and requires a stronger detection. Therefore, a \astroh-like telescope would require $\log F \ga -5$ to be able to measure $R$ with a relative error $<0.1$. However, \nicer\ can achieve that precision for $\log F \ga -6$ which is not much larger than the limit for \athena\ ($\log F \ga -6.5$), whereas \loft\ could provide such precision for $\log F \ga -7$.

\subsection{Comparison to \rxte}
\label{sub:rxte}
The most detailed burst observations to date have been performed with \rxte/PCA \citep[e.g.,][]{1826:galloway04apj}. \rxte\ spectra of bright bursts with the full array of $5$ PCUs exhibit a hint of the \fe\ line \citep{Galloway2008catalog}, and reflection was detected for two superbursts \citep{Ballantyne2004,Keek2014sb2}. For comparison with the other instruments, and to check the reliability of our method, we also create simulations for \rxte/PCA. This instrument consisted of $5$ collimated PCUs that are sensitive in the $2-60$ \kev\ band and have a combined collecting area of $6500\ \mathrm{cm^2}$ \citep{Jahoda2006}. Although only a subset of the 5 PCUs was active during the majority of observations, we employ the response of the full array, which represents the highest quality burst observations. The response is generated using the tool {\sc pcarsp} for the top layer and event mode data with $64$ channels, and the background is estimated with {\sc pcabackest}, where the gain settings from Epoch 5C and background conditions from March 2011 were used as template.

First, to confirm that our method is obtaining the correct uncertainties in the reflection parameters, we include the uncertainties in $\log \xi$ and $R$ for the \rxte\ simulations in Fig.~\ref{fig:reflectuncer}.
We find that the reflection uncertainties are consistent with those measured at the peak of the superbursts near $\log F\simeq -6.0$ for \fouru\ and $\log F\simeq -5.6$ for \eighteen\ (corrected for the different exposure times and the number of active PCUs). Thus, we are confident that the plotted uncertainties in the reflection parameters will be an accurate guide for these future instruments.

\begin{figure}
\centering
\includegraphics[width=0.49\textwidth]{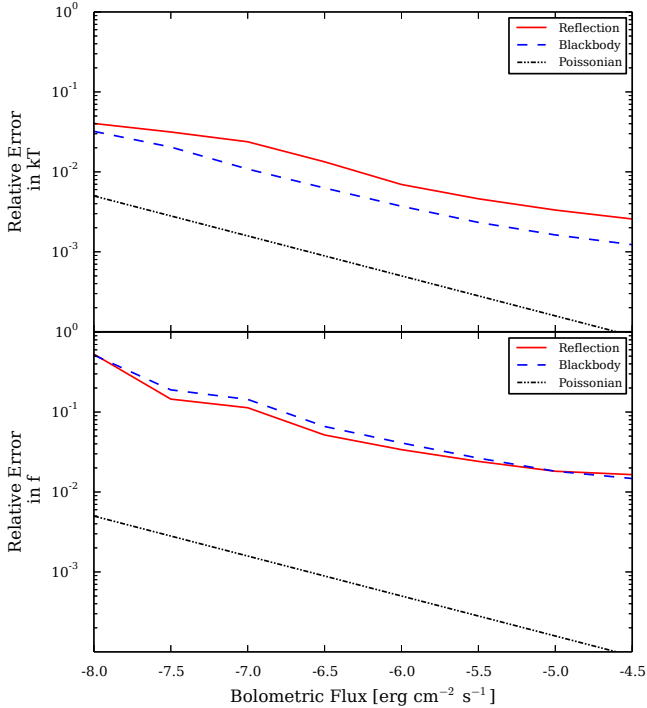}
\caption{Similar to Figure~\ref{fig:uncstack} for \rxte/PCA. For the other instruments the curves are consistent with Poisson noise, especially for large flux values (e.g., Figure~\ref{fig:uncstack}). \rxte, however, exhibits a shallower decline at high flux.}
\label{fig:pca_uncstack}
\end{figure}

Since \rxte\ has been the principle instrument for burst analyses for over a decade, it is useful to show how neglecting the presence of reflection will impact the derived $kT$ and $f$ for \rxte\ observations (Figure~\ref{fig:pca_uncstack}). 
At low flux values, the uncertainties in $kT$ and $f$ are consistent with being scattered around a trend expected for a Poisson distribution. This is similar to our results for \nicer, where at high flux the scatter is reduced and the trend is purely described by Poisson noise (Figure~\ref{fig:uncstack}). For \rxte, however, the trend appears somewhat less steep than a Poisson trend, and it is similar for fits with the blackbody and the reflection models. The same behavior is present for the reflection parameters (Figure~\ref{fig:reflectuncer}). Although at certain flux values \rxte's uncertainties are similar to \nicer's, the scatter is larger. At the largest considered flux, where the statistical quality of the spectra is highest, \rxte\ exhibits the largest relative uncertainties of all considered observatories: a factor $5$ larger for $\log \xi$ and a factor $3$ larger for $R$ compared to \nicer.

Interestingly, the relative uncertainty in $kT$ is systematically larger by a factor $\sim 3$ for the reflection fits compared to the blackbody fits. Given the PCA's band-pass and modest spectral resolution, the broadening of the continuum by reflection may introduce a degeneracy between $kT$ and the reflection parameters. As $kT$ determines the shape of the main spectral component, this also affects the other spectral parameters, including $f$. Furthermore, the main reflection feature observable by the PCA, the \fe\ line, is sampled by just a few energy bins. Measurement of the reflection parameters is highly sensitive to statistical noise in those bins, and this may explain why the \rxte\ simulations exhibit a relatively large scatter in the uncertainties of the reflection parameters. The other instruments sample the reflection signal with a larger number of spectral bins, often including features below $2$ \kev, resulting in a much better behaved measurement of the reflection parameters.

However, the \rxte\ simulations employ the full array of $5$ PCUs, whereas in practice not all were enabled during a particular observation. A typical number of $2$ active PCUs reduces the burst flux by $\Delta \log F\simeq-0.4$. Also, most \rxte\ burst observations have been analyzed at a shorter time resolution of $0.25\ \mathrm{s}$ \citep[e.g.,][]{Galloway2008catalog}, which is equivalent to a further shift in flux of $\Delta \log F\simeq-0.6$. Therefore, in practice most \rxte\ burst analyses are equivalent to the lowest considered flux values, where reflection is not detectable and the behavior of the uncertainty in $kT$ is closest to that of a Poisson distribution. We conclude that neglecting the possibility of reflection in \rxte\ fits of bursts did not greatly impact the measured temperatures, with the deviations being at most $8\%$ in our simulations.

\subsection{Constraining the Reflection Geometry with \loft}
\label{sec:loft}

\begin{figure}
\includegraphics{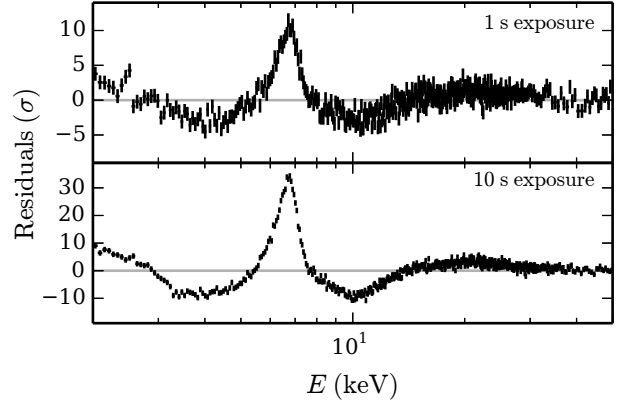}
\protect\caption{\label{fig:loft_residuals}Residuals of blackbody fits to
two \loft\ reflection spectra (in units of the $1\sigma$ uncertainty of the data points). The spectra are simulated for $kT=3\,\mathrm{keV}$,
$\log\xi=3$, and $\log F=-7$. We show a $1\,\mathrm{s}$ (top) and
a $10\,\mathrm{s}$ exposure (bottom). The iron line and edge are
detected at high significance, and allow for the inner disk radius where
the reflection signal originates to be measured.}
\end{figure}
\begin{figure}
\includegraphics{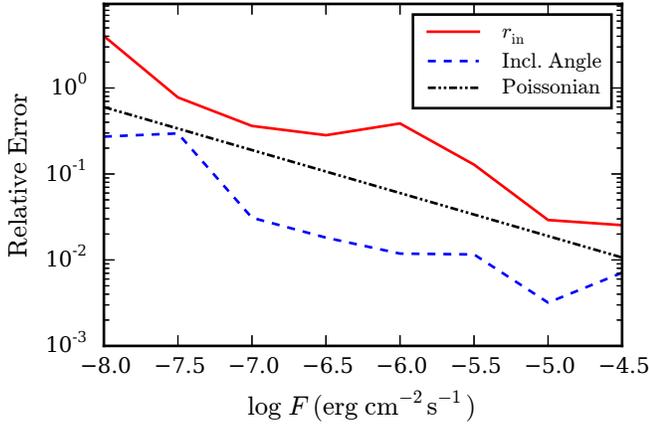}
\protect\caption{\label{fig:loft_rdblur}Relative error in measurements of the inner radius of the reflection location, $r_\mathrm{in}$, and the inclination angle of the disk as a function of the flux, $F$, for \loft. Shown are simulations with $kT=3\,\mathrm{keV}$ and
$\log\xi=3$. The dot-dashed line illustrates the slope of the trend of a Poisson distribution.}
\end{figure}

Contrary to the other considered missions, the band-pass of \loft\ precludes it from detecting the reflection signatures below $2\ \mathrm{keV}$. However, its large collecting area enables such a detailed view of the \fe\ line that \loft\ would provide the most precise measurements of the spectral parameters (Figure~\ref{fig:reflectuncer}).
We illustrate the detectability of the iron line and edge by showing
the residuals of blackbody fits to two reflection spectra with $kT=3\,\mathrm{keV}$
and $\log\xi=3$ (Figure~\ref{fig:loft_residuals}). Because
of the high signal-to-noise ratio, additional parameters can be constrained. We repeat the fits of the reflection model to the simulated \loft\ spectra, leaving the inner radius of the reflection site and the inclination angle of the disk free. The relative errors of the two additional free parameters roughly follow a trend as a function of flux that is expected for a Poisson distribution (Figure~\ref{fig:loft_rdblur}). For a few second exposure
at $\log F=-7$, the inner radius
of the accretion disk can be crudely constrained to several tens of percent, whereas the disk's inclination can be constrained within a few percent.

\subsection{Sources with Low Absorption}
\label{sec:low_abs}

\begin{figure}
\includegraphics{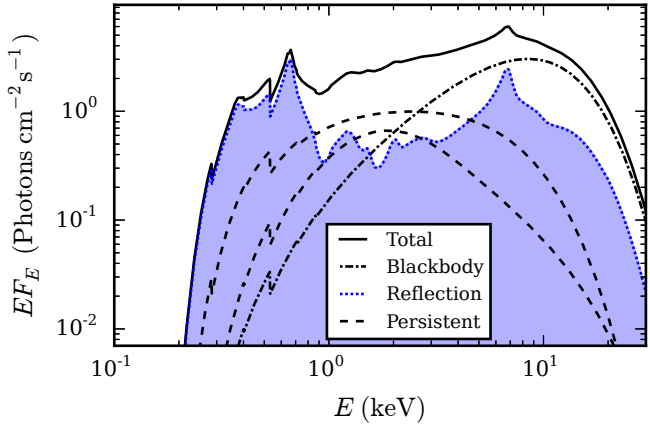}
\protect\caption{\label{fig:model2} Illustration of a spectral model  similar to Figure~\ref{fig:model}, but for a source with lower interstellar absorption: $N_\mathrm{H}=0.1\times 10^{22}\ \mathrm{cm^{-2}}$. Burst reflection (shaded area) dominates the X-ray spectrum below $\sim 1\ \mathrm{keV}$.}
\end{figure}

\begin{figure}
\centering
\includegraphics[width=0.49\textwidth]{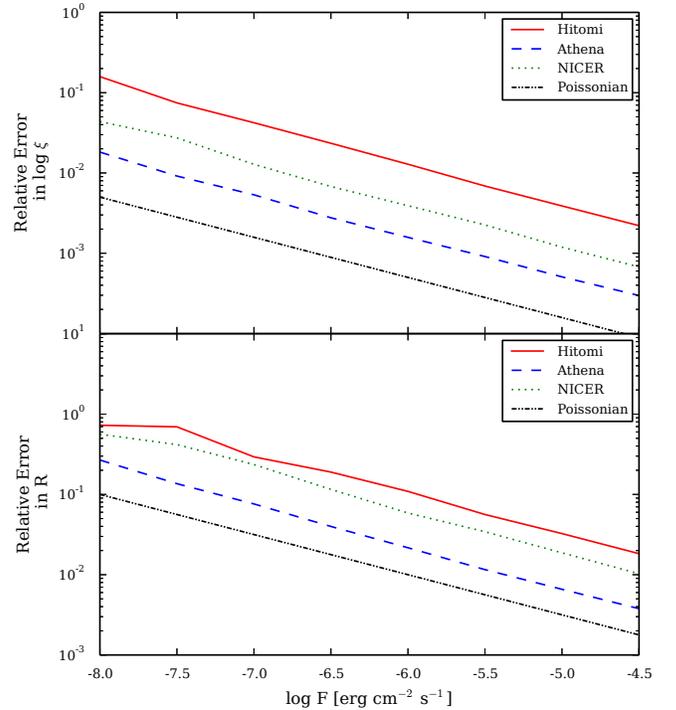}
\caption{Similar to Figure \ref{fig:reflectuncer}, for simulations with low absorption ($N_\mathrm{H}=0.12\times 10^{22}\ \mathrm{cm^{-2}}$) of the three instruments that are sensitive below $2$~\kev.}
\label{fig:reflectuncer_abs}
\end{figure}

\begin{figure}
\includegraphics{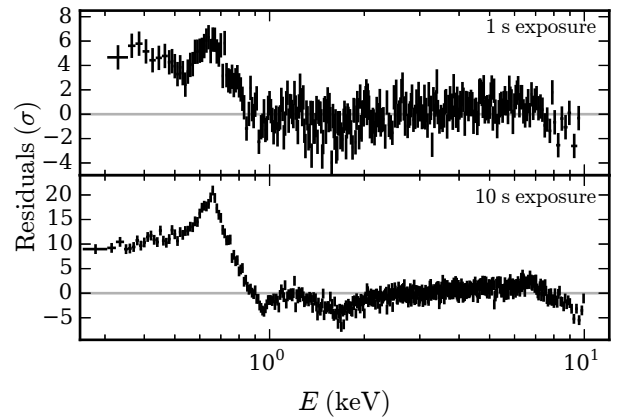}
\protect\caption{\label{fig:nicer_residuals_lowNH}Similar to Figure~\ref{fig:nicer_residuals} for a small absorption column of $N_\mathrm{H}=0.12\times 10^{22}\mathrm{cm^{-2}}$, which allows the reflection features below $\sim 1$\,\kev\ to be detected at high significance by \nicer.}
\end{figure}

We have employed a relatively high absorption column that is typical for sources in the Galactic plane \citep[$N_\mathrm{H}=0.891\times 10^{22}\ \mathrm{cm^{-2}}$ for \oheight;][]{Keek2008}, and therefore the predicted soft features are largely erased by interstellar absorption (Figure~\ref{fig:model}). A select few known bursting sources have a smaller absorption column of $N_\mathrm{H} \approx 0.1\times 10^{22}\ \mathrm{cm^{-2}}$. This allows for the detection of emission features in the spectrum below $\sim 1\ \mathrm{keV}$ (Figure~\ref{fig:model2}), where burst reflection dominates the soft part of the spectrum. For \astroh, \nicer, and \athena, which cover the soft band, we redo the simulations using $N_\mathrm{H}=0.12\times 10^{22}\ \mathrm{cm^{-2}}$. This is the absorption column measured for SAX~J1808.4-3658 \citep{Wang2001}. The rest of the spectral model is the same as before. For all three instruments the uncertainties in the reflection parameters are reduced for a given flux compared to the simulation with a larger $N_\mathrm{H}$ value (Figure~\ref{fig:reflectuncer_abs}). For example, \nicer\ reaches a precision for $\log \xi$ of $\sim 2\%$ in $1\ \mathrm{s}$ and measurements of the reflection fraction are improved as well. This is due to the large number of soft photons below $1$ \kev\ that constitute the most prominent part of the reflection signal in this case (Figure~\ref{fig:nicer_residuals_lowNH}). Furthermore, the scatter around the Poisson trend at low flux is reduced with respect to the simulations with higher absorption (Figure~\ref{fig:reflectuncer}).

\subsection{Limitations on Burst Observations due to Pile up}
\label{sec:Pileup}
A common issue for X-ray detectors is ``pile up.'' If multiple photons are detected by the same or neighboring pixels within one read-out cycle, they may be recorded as a single photon with an energy that is the sum of the photon energies. This reduces the observed count rates and distorts spectra. For \astroh\ SXI in the fastest read-out mode, a burst spectrum with $kT=3.0$~keV and $\log\xi=3.0$ is expected to be piled up at a level of 10\% for fluxes in excess of $\gtrsim 2\times 10^{-9}\ \mathrm{erg\ s^{-1}cm^{-2}}$ \citep{Tsunemi2013AstrohSXI}, which covers the entire flux range that we considered in our simulations. Possibly, this issue could have been mitigated by a faster read-out mode, similar to the ``burst'' mode of {\it XMM-Newton}'s EPIC pn instrument. Such a mode was not planned for \astroh\ \citep{Tsunemi2013AstrohSXI}.

Operation of \athena\ WFI will include a small window mode that facilitates high count rates of up to $7.8\times 10^3\ \mathrm{c\ s^{-1}}$ \citep{Meidinger2014AthenaWFI}. X-ray bursts may, however, reach this limit already for a flux of $\sim 2\times 10^{-8}\ \mathrm{erg\ s^{-1}cm^{-2}}$. This flux is close to the minimum flux required for detecting reflection. For bursts with fluxes below this limit, an integration time of several seconds may be sufficient to collect spectra with detectable reflection features. Conversely, for brighter bursts, the usual strategy of dealing with pile up may be applied: the center of the point spread function where pile up is present, is excluded from the spectra. This reduces the statistical quality of the spectra. However, similar to the weaker bursts, reflection features may still be detected. Moreover, as the launch date of \athena\ is still far in the future, technical solutions could be found to increase the count rate limit, such as a fast read-out mode.

The detectors on board \nicer\ are of a different type, and were specifically designed to handle very high count rates. For  $\log F=-7.0$, $kT=3.0\ \mathrm{keV}$, and $\log\xi=3.0$, our spectral model predicts a count rate of $7.2 \times 10^3\ \mathrm{c\ s^{-1}}$. The signal will be divided among 56 detectors, allowing \nicer\ to easily handle such large rates. Similarly, dead time (the time a detector is inactive when an event is being processed) is negligible for typical bright bursts. Observations with \nicer\ can, therefore, be used to the full extent to detect burst reflection. The same is true for \loft, which likewise was designed to handle high throughput.

\section{Discussion}
\label{sect:discuss}

\subsection{Detectability of Burst Reflection}

Detecting the presence of accretion disk reflection in X-ray burst spectra is challenging because of the short duration of the bursts and the mixture of multiple emission components all contributing to the observed source. We showed that the detectability of reflection features depends only mildly on the blackbody temperature $kT$, with lower $kT$ being easier to detect as more of the emission falls in the band-passes of most of the modeled instruments. In contrast, the reflection features are significantly easier to detect for ionization parameters $\log \xi \approx 2.5-3$ (Figures~\ref{fig:chicompare} and~\ref{fig:Ftestcompare}) because of the large equivalent width of the \fe\ line and a strong soft excess. The larger values of the ionization parameter will likely occur near the onset of any burst when the flux on the disk is largest \citep{Ballantyne2004,Keek2014sb2}. Thus, searches for reflection features in burst spectra will have the best chance for success with observations that are as close to the start of the burst as possible. 

We investigated the detection of burst reflection with four instruments. For bursts with a flux of $\log F \ga -7.5$, \nicer\ will significantly detect the presence of disk reflection in burst spectra, and will provide tight constraints on reflection parameters for 1~s exposures at $\log F \ga -6$. Our simulations all employ exposure times of 1~s, but the results are applicable to longer exposures for proportionally lower fluxes (such that the total number of counts is the same). Therefore, for spectra originating from a bright burst of sufficient duration, \nicer\ will be able to also track how the features change in time in response to the burst. 

Unfortunately, the relatively low effective area of a \astroh-like telescope means that although it will be able to statistically detect the reflection signal for bursts with $\log F \ga -7$, it will be unable to constrain reflection parameters for any realistic burst or superburst. Aside from the instrument that we consider (the SXI), \astroh\ also hosted the Soft X-ray Spectrometer (SXS). The SXI and SXS had overlapping  band-passes and similar limitations on the photon rate that can be processed. Including the SXS would have doubled the collecting area, and this reduces the flux requirements from our simulations by a factor $2$ ($\Delta \log F \simeq 0.3$). This is insufficient to substantially improve the detectability of burst reflection with \astroh.

In the next decade, \athena\ will provide an order-of-magnitude improvement over \nicer\ in detecting reflection features from bursts. With \athena\ it may be possible to directly constrain reflection features in a 1~s exposure of bright bursts, opening up the possibility of tracing the evolution of the reflection parameters during a burst in detail. An important issue is for an observing mode to be devised for the WFI to handle the large count rates expected from the brightest bursts in order to avoid pile up. Similar to \astroh, \athena\ will also include a spectrometer that has a similar energy band as the WFI, and that can be employed to increase the collected number of photons by approximately a factor $2$.

A \loft-like mission with a $\sim 8.5$~m$^{2}$ collecting area would revolutionize this field and be able to perform accretion disk tomography using the \fe\ line from bright Type I X-ray bursts. This would provide detailed information on the time evolution of the accretion environment of neutron stars under the influence of strong X-ray irradiation.

\subsection{The Soft X-Ray Band and the Influence of Absorption}

The broad band-passes of the new and upcoming instruments present new opportunities to study the disk reflection signal. Models of the reflection features predict a soft excess in the reflection spectrum due to both a bremsstrahlung continuum and recombination lines \citep{Ballantyne2004models}. Unfortunately, for most known bursting sources the interstellar absorption is large, and the signal in the soft band is substantially reduced. Still, the remaining signal is sufficient to enable the measurement of the reflection parameters during short exposures with \nicer\ and \athena. Conversely, \rxte\ does not cover the soft band, and only detects the \fe\ line and absorption edge. During short exposures, we find that \rxte's reflection detection is strongly impacted by statistical noise in the few spectral bins that cover these features, which explains why reflection was only observed clearly in longer exposures for two superbursts. Therefore, coverage of the soft band is crucial for \nicer\ and \athena\ to measure the reflection properties, even in the presence of strong absorption.

The recombination lines below $1$~\kev\ carry information on the disk composition and density \citep{Ballantyne2004models}. For a few sources the interstellar absorption is low enough to observe these features in detail: only $2.6\%$ of known bursts originate from sources with $N_\mathrm{H}\le 0.2\times 10^{22}\ \mathrm{cm^{-2}}$ \citep[e.g.,][]{Cornelisse2003,Galloway2008catalog}.
An example of a source with a small absorption column is SAX~J1808.4-3658 with $N_\mathrm{H}=0.12\times 10^{22}\ \mathrm{cm^{-2}}$ \citep{Wang2001}. A {\it Chandra} observation of a burst from this source (with contemporaneous \rxte/PCA coverage) exhibits a soft excess with respect to a blackbody model, which is well fit with a model of reflection off a highly ionized accretion disk \citep{Zand2013}. However, the quality of the spectra was insufficient to distinguish burst reflection from, e.g., increased persistent emission. We find that observations with future instruments that are sensitive to the soft X-ray band will provide the strongest constraints on the reflection parameters. 

The recently launched {\it ASTROSAT} \citep{Singh2014Astrosat} has an array of Large Area Xenon Proportional Counters (LAXPC) similar to the PCA, as well as a Soft X-ray Telescope (SXT). The effective area at $6.4$~keV of LAXPC is similar to that of \rxte/PCA. The SXT extends {\it ASTROSAT}'s coverage to lower energies. Its effective area at $1$~keV is, however, only $120\ \mathrm{cm^2}$, which is smaller than {\it Chandra}. We, therefore, expect that {\it ASTROSAT} will only observe reflection in long bursts, similar to \rxte.

\subsection{Applications of Burst Reflection}

In the same way as X-ray reflection has proved to be invaluable for studying the accretion processes around black holes, we expect burst reflection to provide invaluable insight into accretion physics in the vicinity of neutron stars. Here we discuss several important applications that will be enabled by observations of burst reflection.

\subsubsection{Mass and Radius Measurements of Neutron Stars}

In recent years X-ray bursts have been employed to measure the masses and radii of neutron stars in order to constrain the equation of state of dense matter \citep[e.g.,][]{Paradijs1979,Ozel2006,Steiner2010,Guver2012,Poutanen2014}.\cite{} X-ray bursts provide the rare opportunity to determine both quantities simultaneously \citep[e.g.,][]{Lattimer2007}. A crucial issue is the accurate interpretation of burst spectra, as deviations from a blackbody introduce systematic uncertainties. For example, free-free and Compton scattering in the neutron star atmosphere introduce subtle changes to the spectral shape. Detailed spectral models have been created for neutron star atmospheres during X-ray bursts \citep[e.g.,][]{Suleimanov2010}, and have been successfully fit to certain burst spectra \citep{Suleimanov2011}. For other observations, however, the atmosphere models do not reproduce the expected behavior \citep{Kajava2014}, especially in the high-flux soft persistent state. In this spectral state, the accretion disk is thought to extend closest to the neutron star \citep{Done2007review}, such that the reflection signal is maximal. The reflection spectrum is also a reprocessed blackbody, which deviates more strongly from a blackbody (Figure~\ref{fig:model2}). Therefore, a substantial contribution of reflection to the burst spectrum may explain those observations that are not well described by burst atmosphere models.

By detecting burst reflection, its contribution to the spectrum can be quantified. This provides guidance for mass-radius measurements to select those bursts that are least ``contaminated.'' Moreover, if the reflection parameters can be sufficiently well determined, the properties of the burst atmosphere and the reflection components may be constrained simultaneously. This will also be important for studies that use the shape of burst light curves to constrain nuclear reactions among short-lived proton-rich isotopes \citep[e.g.,][]{Fisker2006,Cyburt2010,Parikh2013review}.

\subsubsection{Evolution of the Accretion Environment}

For X-ray reflection of coronal emission in AGN and compact binaries, the reflection signal is observed to evolve over time \citep[e.g.,][]{Ballantyne2011,Keek2016Mrk335}. These observations simultaneously probe changes in the corona and in the disk, making it challenging to uncover the evolution of the individual regions. For burst reflection the situation is less complicated, as the neutron star's geometry remains largely constant, with the possible exception of a brief well discernible period of radius expansion. Furthermore, the time evolution of the burst emission from the neutron star can be successfully modeled in great detail \citep{Heger2007,Zand2009}. Therefore, the evolution of the reflection parameters during X-ray bursts predominantly probes changes in the accretion disk. 

X-ray bursts provide a repeating experiment to investigate the response of accretion disks to sudden strong irradiation. During the two superbursts that at present provided the only clear detections of burst reflection, the accretion disks were found to be strongly ionized at the onset of the events, and the ionization parameter was reduced during the tail of the bursts \citep{Ballantyne2004,Keek2014sb2}. For the 1999 superburst from \eighteen\ the reflection signal was sufficiently strong to trace the location of reflection: it initially was located further from the neutron star before returning to smaller radii in the tail, suggesting that this superburst initially disrupted the inner disk. In this paper we find that with future instruments, this type of analysis can be performed also for short bursts, which are detected at a thousand times higher rate than superbursts.

The reflection fraction, $R$, is an important parameter for tracing changes in the geometry of the disk. A flat disk produces a value of $R=0.5$ \citep{lapidus85mnras,fujimoto88apj}. For the two mentioned superbursts, larger values may have been present \citep{Ballantyne2004,Keek2014sb2,Keek2015sb}, which could indicate that the inner disk was puffed up due to the burst irradiation \citep{He2015}. X-ray heating of the disk may, however, not be the only important process during bursts. The generation of winds or the inflow of material by Poynting-Robertson drag could play a role as well \citep{Ballantyne2005}. Further theoretical studies are required to investigate this in more detail, and we expect that burst observations with future instruments will provide constraints for such models.

\section{Conclusions}
\label{sect:conclusions}

Reflection spectroscopy during Type I X-ray bursts holds the promise to open a new avenue for studying accretion physics in the vicinity of neutron stars, by revealing the response of accretion disks to sudden strong irradiation. Furthermore, burst reflection can act as a quantitative measure of which bursts are most suitable for mass-radius determination. Previously, reflection has only been detected during two superbursts. We investigate the detectability of burst reflection using a large set of simulated spectra that are representative of most known bursts. Taking into account the instrumental energy response, effective area, and ability to handle large count rates, we find that future X-ray observatories will be able to detect reflection during the frequent short Type I bursts.
Considering all these factors leads us to conclude that \nicer\ provides an excellent opportunity to study the interaction between X-ray bursts and the surrounding accretion disk that will not be significantly surpassed until the launch of \athena. Further in the future, an observatory with a large collecting area similar to the \loft\ design would enable studying the accretion processes in unprecedented detail.

\acknowledgments{
The authors thank R.E.~Rutledge for encouraging to write this paper and T.E.~Strohmayer for helpful comments. 
LK is supported by NASA under award number NNG06EO90A. LK thanks the
International Space Science Institute in Bern, Switzerland for hosting an
International Team on X-ray bursts.}

\bibliographystyle{apj}
\bibliography{references}

\end{document}